\documentclass{jpsj3}
\usepackage{txfonts}
\usepackage{amssymb}
\usepackage{graphicx}
\usepackage{amsmath}
\usepackage{amsfonts}
\usepackage{amssymb}
\usepackage{enumitem}  
\usepackage{bm}

\title{Exchange Enhancement of the Electron-Phonon Interaction: the
  Case of Weakly Doped Two-Dimensional Multivalley Semiconductors}

\author{Betul Pamuk$^1$, Paolo Zoccante$^2$, Jacopo Baima,$^2$, Francesco Mauri$^{3,4}$ and Matteo Calandra$^2$}
\inst{$^1$School of Applied and Engineering Physics, Cornell University, Ithaca, New York 14853, USA\\
$^2$Sorbonne Universit\'es, UPMC, CNRS-UMR 7588, Institut des NanoSciences de Paris, F-75005, Paris, France \\
$^3$ Dipartimento di Fisica, Universit\'a di Roma La Sapienza, Piazzale
Aldo Moro 5, I-00185 Roma, Italy\\
$^4$ Graphene Labs, Fondazione Istituto Italiano di Tecnologia, Italy} 

\abst{The effect of the exchange interaction on the vibrational
  properties and on the electron-phonon coupling were investigated in
  several recent works. In most of the case, exchange tends to
  enhance the electron-phonon interaction, although the motivations
for such behaviour are not completely understood. 
Here we consider the class of weakly doped two-dimensional multivalley
semiconductors and we demonstrate that a more global picture emerges.
In particular we show that in these systems, at low enough doping, even a moderate electron-electron
interaction enhances the response to any perturbation inducing a
valley polarization.
If the valley
polarization is due to the electron-phonon coupling, the electron-electron interaction results in an
enhancement of the superconducting critical temperature.
We demonstrate the applicability of the theory by performing 
random phase approximation and first principles calculations in
transition metal chloronitrides. We find that exchange is responsible
for the enhancement of the superconducting critical temperature in
Li$_x$ZrNCl
and that much larger T$_c$s could be obtained in intercalated HfNCl if the
synthesis of cleaner
samples could remove the Anderson insulating state competing
with superconductivity.}

\begin{document}
\maketitle

\section{Introduction}

Calculations of superconducting properties of materials require the
knowledge of the electronic structure, the vibrational spectrum and a
good description of the electron-phonon interaction. In
state-of-the-art electronic structure approaches, density functional perturbation
theory \cite{LRBaroni,GonzePh} with local functionals (local density approximation or
generalized gradient approximation) is used.
However, recently it has been shown that 
in several case the treatment of the electron-electron interaction in this framework is 
unsatisfactory and it becomes necessary to include a certain amount
of exact exchange in the calculations. 
This is often the case in 2D or layered systems with
weak Van der Waals out-of-plane interactions such as
Graphene \cite{Attaccalite,Attaccalite2}, transition metal chloronitrides
\cite{Paolo2015,Pamuk2016,Pamuk2017,Kotliar2013} or transition metal dichalcogenides
\cite{Hellgren2017}. However, enhancement of the electron-phonon
coupling due to the electron-electron interaction has also been shown
to occur in diamond\cite{AntoniusPRL2014} and in
molecules
like C$_{60}$ in gas phase\cite{  FaberPhysRevB.84.155104,JanssenPhysRevB.81.073106}.
Despite all these recent calculations, there is not a clear
understanding of when and why in a given system the exchange
interaction or, more generally, the electron-electron interaction has
important consequences for the electron-phonon coupling.

In the special case of a weakly doped multivalley two dimensional
semiconductor, a more general picture of why the
electron-electron
interaction has important consequences in the renormalization of the
electron-phonon coupling and of vibrational properties has been
developed\cite{Paolo2015,Pamuk2016,Pamuk2017}. It relies
on the general fact that at weak doping (few
electrons/holes per valley) 
manybody effects enhance the response to any
perturbation inducing a valley polarization. A physical realization of
a multivalley 2D electron gas are Al/AS quantum wells. It has been 
shown that in this case a strain deformation acts as a {\it pseudo} 
magnetic field inducing a valley polarization , as it was shown in Ref. \cite{Gunawan}.
Another form of a pseudo-magnetic field could be an intervalley phonon, namely a phonon with a
momentum coupling two different valleys, inducing a change in the
occupation of the valleys. In this case the pseudo magnetic field is
simply the modulus of the deformation potential associated to this
vibration. In all case, the response to the pseudo magnetic field is
enhanced by the electron-electron interaction.

This effect can be justified by using a model Hamiltonian
with SU$(2g_v)$ valley-spin symmetry\cite{Ando1982,Marchi2009,DasSarma2009},
with $g_v$ being the number of valleys. This symmetry enforces the
constrain that the interacting  spin and valleys susceptibilities must
be identical. As the spin susceptibility is enhanced by manybody
effects
in the low doping limit\cite{Marchi2009}, the same must occurs to the valley
susceptibility.
 In real systems, the SU$(2g_v)$ symmetry holds for an
isotropic mass tensor and at low doping. Indeed in this case, 
the Fermi momentum $\kappa_F$,
as measured from the valley bottom, is much
smaller than the valley separation and, consequently,
the intravalley electron-electron interaction dominates over the
intervalley one. This will be shown in more details in
sec. \ref{secmodel}.

A class of systems where these assumptions hold are the transition metal
chloronitrides (examples are  HfNCl and ZrNCl) that are semiconductors
crystallizing in hexagonal structures and having almost perfect
parabolic conduction bands at the high symmetry points ${\bf K}$ and
$2{\bf K}$. The interlayer interaction is extremely weak so that
intercalation leads to an almost perfect realization of a 2D
multivalley electron-gas\cite{Heid2005,Paolo2015}.
Intercalation with alkali metals or alkaline earths
induces a metal insulator transition at doping
ranging from $x=0.05$ to $x=0.14$, depending on the system.
In the metallic state and at low temperature a superconducting
state occurs with T$_c$ that can be as high as $25$K
\cite{Yamanaka1996,Yamanaka1998,Taguchi2006,Takano2008}.
The behaviour of T$_c$ versus doping is very peculiar as T$_c$
{\it decreases} with {\it increasing} doping, in contrast with their
doping independent density of states (2D parabolic bands)\cite{Taguchi2006}.
We will show in the sections below that this anomalous behaviour is
due to the fact that the exchange interaction enhances the electron-phonon
coupling in the low doping limit. 

   \section{Spin and valley susceptibilities in a 2D multivalley electron gas\label{secmodel}}

We consider a two dimensional multivalley electron gas composed by an isolated band partially filled with electrons, namely
 \begin{equation}
H_0=\sum_{{\bm \kappa} v \sigma}\frac{\hbar^2 \kappa^2}{2 m^*}
c^{\dagger}_{{\bm \kappa} v\sigma} c_{{\bm \kappa} v\sigma}
\label{Eq:H_0}
\end{equation}
where $v=1,...,g_v$ is a valley index with $g_v$ being the valley degeneracy, $\sigma=\pm$ is a spin index and $c$,$c^{\dagger}$ are
fermion creation and destruction operators. The vector $\kappa$ is
measured from the bottom of each valley. In two dimensions,  the Fourier transform of
the electron-electron interaction within this band
reads
\begin{equation}
v(q)=\frac{2 \pi e^2}{\epsilon_M q}
\end{equation}
where ${\bf q}$ is the exchanged momentum between the two interacting electrons,
and the effect of the screening of other (empty) conduction and (filled)
valence bands is included via the environmental
dielectric constant $\epsilon_M$. 
The electron-electron interaction
has two contributions:
\begin{enumerate}[label=(\roman*)]
\item  The intravalley scattering with $q\sim\kappa_F$ (
$\kappa_F$ being the Fermi momentum measured from the valley bottom).
\item  The intervalley scattering with $q\sim|{\bf K}-{\bf
    K}^{\prime}|=|{\bf K}|$
where ${\bf K}$ and ${\bf K^{\prime}}=2{\bf K}$ are the positions of the valley
bottoms in the Brillouin zone.
\end{enumerate}
The intravalley contribution conserves the  valley index of the
electrons while the intervalley contribution does not.

In the low doping limit, namely for $\kappa_F\ll |{\bf K}-{\bf K}^{\prime}|$,
because of the divergence of the Coulomb repulsion for $q\to 0$, the
intravalley scattering is dominant and the intervalley scattering can be neglected.
Under this hypothesis, the valley and spin index are conserved by the
Coulomb interaction.
The valley index can then be treated as a pseudospin and
the manybody Hamiltonian has exact SU$(2 g_v)$
spin and valley symmetry, namely (see e. g. Eq. 3.35 of 
Ref. \cite{Ando1982})
\begin{eqnarray}
H&=&\sum_{{\bm \kappa} v \sigma}\frac{\hbar^2 \kappa^2}{2 m^*}
c^{\dagger}_{{\bm \kappa} v\sigma} c_{{\bm \kappa} v\sigma} +\nonumber \\
&+&\sum_{{\bm \kappa} v\sigma}
\sum_{{\bm \kappa^{\prime}} v^{\prime} \sigma^{\prime}}
\sum_{\bm q} v(q) 
c^{\dagger}_{{\bm \kappa} v\sigma} 
c^{\dagger}_{{\bm \kappa^{\prime}} v^{\prime} \sigma^{\prime}}  
 c_{{\bm \kappa^{\prime}}-{\bm q} v^{\prime}\sigma^{\prime}} 
 c_{{\bm \kappa}+{\bm q} v\sigma} 
\label{Eq:H_ando}
\end{eqnarray}
The Hamiltonian in Eq. \ref{Eq:H_ando} holds as long as
(i) the screening of the other bands can be included in the environmental
dielectric constant, (ii) intervalley scattering can be neglected. If
these two conditions are satisfied, then it holds regardless of the
number of valleys and of their position in the Brillouin zone.

As the Hamiltonian in Eq. \ref{Eq:H_ando} has exact SU$(2 g_v)$
spin and valley symmetry, it follows that:
\begin{equation}
\chi_v = \chi_s
\label{eqsymchis}
\end{equation}
where the relation holds both for the interacting and non interacting susceptibilities.
As a consequence, the knowledge of the interacting spin susceptibility is
equivalent to the determination of the interacting valley
susceptibility. 
\begin{figure}[h]
\vspace{2.5 cm}
\hspace{-1.0cm}\includegraphics[width=0.6\columnwidth]{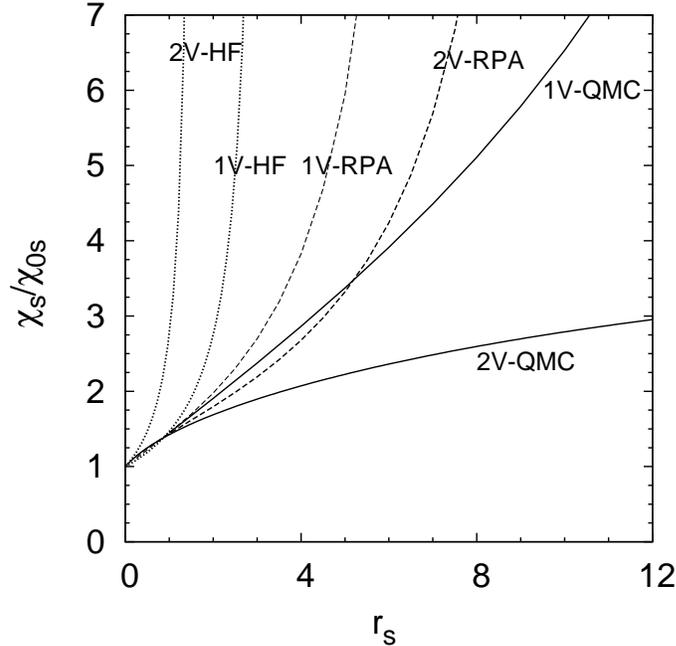}
\vspace{-2.5 cm}
\caption{Ratio between the interacting susceptibility calculated 
with several approximations and the bare susceptibility $\chi_{0s}$
for a two dimensional  
1 (1V) and 2 (2V) valley electron gas as a function of 
$r_s=\frac{m^{*} e^2}{\epsilon_M \hbar^2 \sqrt{\pi n}}$ ( HF=
Hartree-Fock, RPA=Random Phase Approximation, QMC=Quantum Monte Carlo).
Figure adapted with permission from Ref. \cite{Marchi2009}. 
Copyrighted by the American Physical Society }
\label{figs_chi_marchi}
\end{figure} 
In the absence of electron-electron interaction the non interacting
spin susceptibility is $\chi_{0s}=\mu_S N(0)=g_v m^*/\pi \hbar^2$
independent of doping and of the electron-gas parameter
$r_s=\frac{m^{*} e^2}{\epsilon_M \hbar^2 \sqrt{\pi n}}$, with $n$
being
the electron-density .

The interacting spin susceptibility for a two dimensional  two
valley electron-gas has been calculated in detail
Ref. \cite{Marchi2009}. The results of this calculation are reported in
Fig. \ref{figs_chi_marchi}. As it can be seen the Hartree Fock (HF)
approximation leads to strongly divergent susceptibilities already at
fairly low values of $r_s$, with the two valley case more divergent
than the one valley case. This means that the exchange interaction
leads to a magnetic solution.
The Random Phase Approximation (RPA) removes 
the divergence, reduces the magnitude of the interacting spin 
susceptibility with respect to the 
HF case and leads to a smaller $\chi_s$ for the 2 valley
electron-gas, in contrast to the HF result. 
The inclusion of correlation effects with the most accurate Quantum
Monte Carlo (QMC) technique shows that correlation tends to reduce
the exchange-induced enhancement in $\chi_S/\chi_{0s}$.
Finally, it can be seen that for a 2 valley electron gas and $r_s<2$,
 the RPA is an excellent approximation to the QMC results.
As of Eq. \ref{eqsymchis}, the same enhancement must occur in 
$\chi_v$. 
\begin{figure}[h]
\vspace{0.5 cm}
\centerline{\includegraphics[width=0.6\columnwidth]{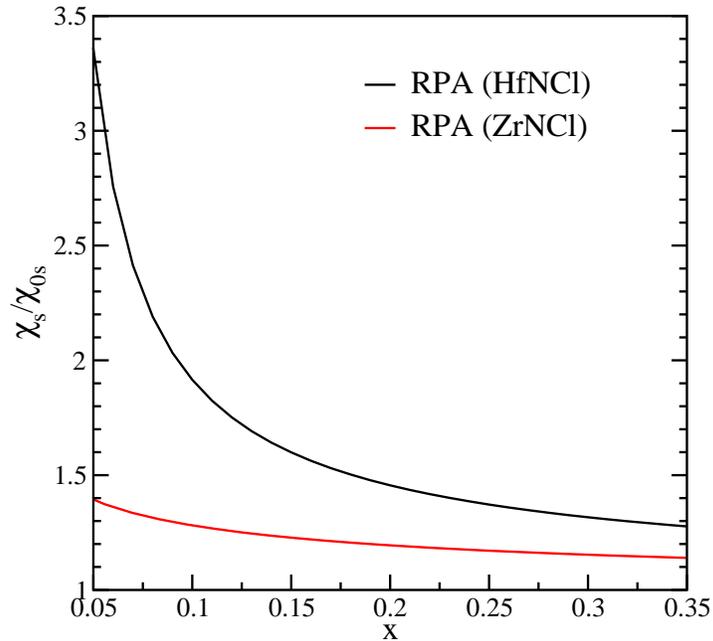}}
\caption{(Color online) Ratio between the interacting susceptibility calculated in the Random  
  phase approximation and the bare susceptibility $\chi_{0s}$ for a 2D  
2 valley electron gas using relevant parameters of Li$_x$ZrNCl and Li$_x$HfNCl.  
 For Li$_x$ZrNCl  
we use $\epsilon_M=5.59$ and $m^{*}=0.57 m_e$, while for Li$_x$HfNCl we used 
$\epsilon_M=4.93$ and $m^{*}=0.615 m_e$ where $m_e$ is the electron  
mass.  
The larger divergence of the susceptibility in intercalated HfNCl is  
mostly due to the smaller environmental dielectric constant. }
\label{figs_chi_enhancement}
\end{figure} 
In Fig. \ref{figs_chi_enhancement} we plot the interacting spin susceptibility
for a two valley 2D electron-gas within the RPA 
 (see  supplemental materials of \cite{Paolo2015} for more details), 
using  relevant parameters 
for ZrNCl and HfNCl compounds\cite{Paolo2015,Pamuk2016,Pamuk2017}.
As it can be seen the spin susceptibility is enhanced at low doping
and large $r_s=\frac{m^{*} e^2}{\epsilon_M \hbar^2 \sqrt{\pi n}}$ with
respect to the doping-independent bare (non interacting) susceptibility. The enhancement
is entirely due due to the electron-electron interaction. The enhancement
of the spin susceptibility at low doping has been measured in
Li$_x$ZrNCl\cite{Taguchi2006}and it is in perfect agreement with RPA\cite{Paolo2015}.

\section{Electronic and vibrational properties of intercalated
  (Zr,Hf)NCl }

\subsection{Crystal structure}

The primitive unit cell of intercalated (Zr,Hf)NCl has rhombohedral structure (space
group $R\bar{3}m$, number $166$) with 2 formula units per unit cell.
It can also be constructed by a conventional cell of hexagonal structure
with 6 formula units per cell with ABC stacking.
It has been shown that the weak
interlayer interaction \cite{Kasahara2010,Heid2005,Takano2011,Botana2014, Paolo2015},
makes the stacking order negligible so that simulations with
an hexagonal structure and AAA stacking (space group number 164)
an  2 formula units unit cell leads to the same results.

\subsection{Electronic structure}

\begin{figure}[h]
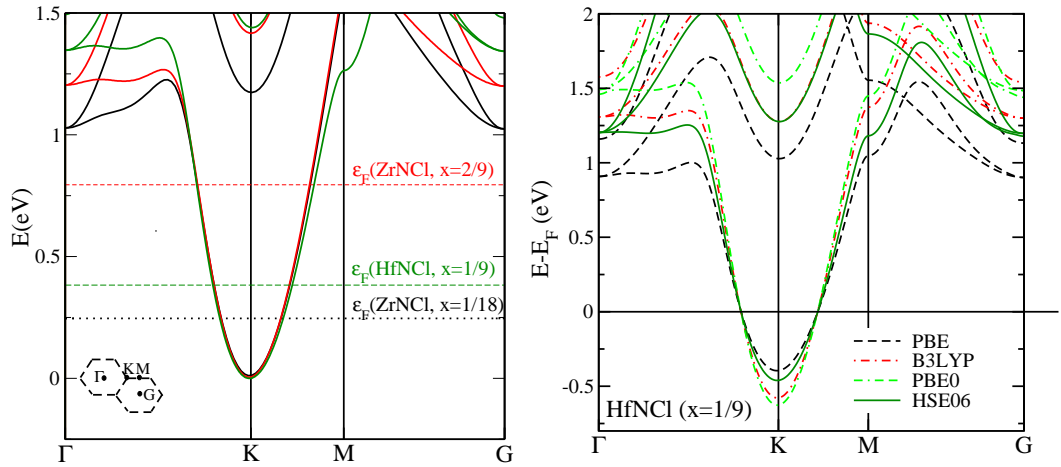

\vspace{1 cm}
\centerline{\includegraphics[width=0.43\columnwidth]{Figure_bands_only.eps}
\includegraphics[width=0.452\columnwidth]{bands_HfNCl_functional.eps}}
\caption{(Color online) Top:electronic structure of intercalated (Zr,Hf)NCl for 
  several doping (generalized gradiend approximation within the PBE parametrization\cite{PBE}). 
Bottom:effect of the exchange interaction on the electronic structure 
of doped HfNCl at $x=1/9$. }
\label{fig_el_struc}
\end{figure}

\begin{figure}[h]
\centerline{\includegraphics[width=0.5\columnwidth]{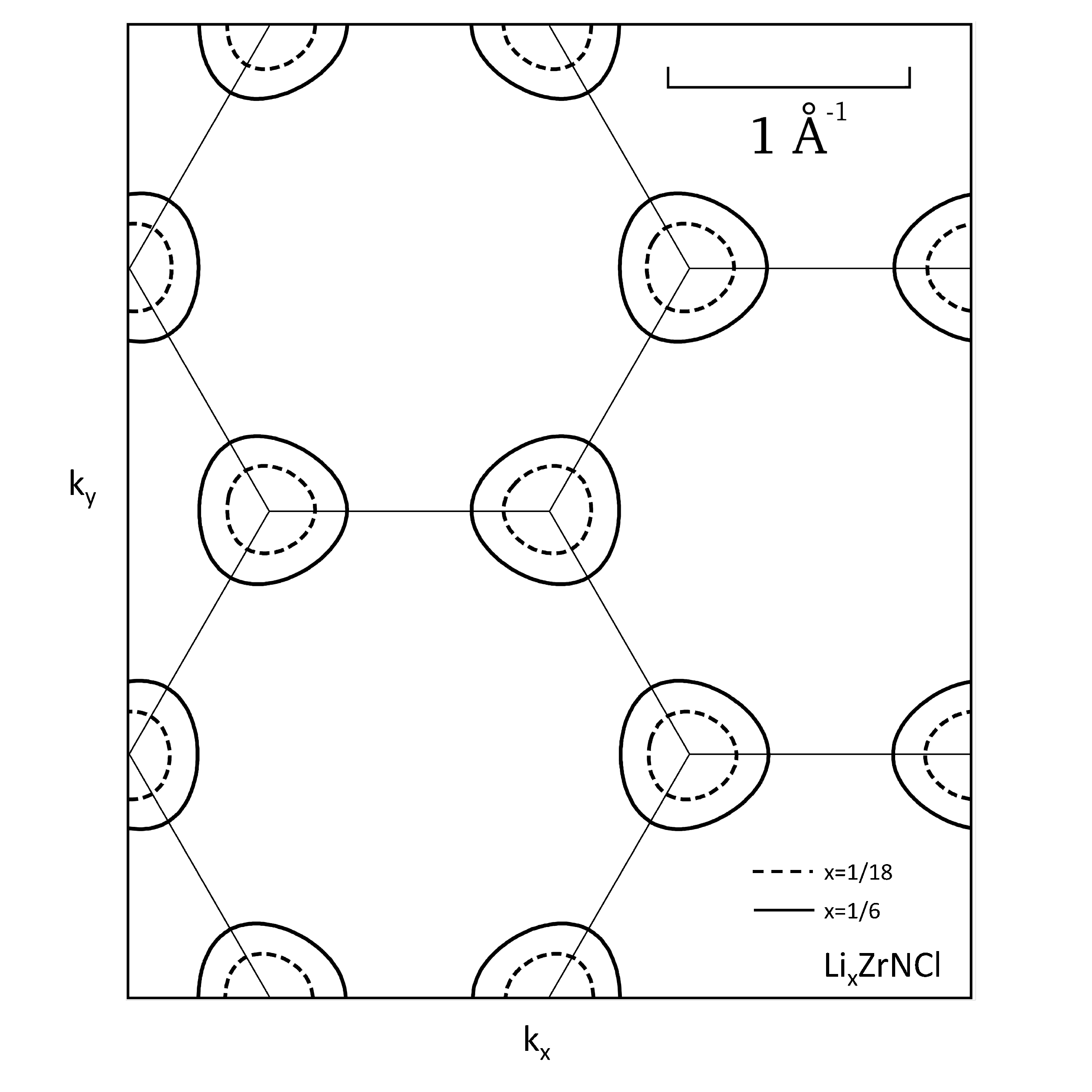}}
\caption{In plane projection ($k_z=0$) of the Li$_x$ZrNCl Fermi surface as a function of doping.}
\label{fig_FS}
\end{figure}

The electronic structure of intercalated (Zr,Hf)NCl using local
functionals has been
calculated by many
authors\cite{Heid2005,Weht1999,Akashi2012,Kotliar2013,Paolo2015,Pamuk2016,Pamuk2017}. 
The general
agreement is that the charge transfer from alkali metals is
complete and it acts as a rigid doping, so that it is appropriate
to simulate these systems by using a uniform background doping.
This has been explicitly verified both for the electronic and
vibrational properties \cite{Heid2005}. Under this assumption, 
the electronic structure calculated using the generalized gradient
approximation
(GGA)  approximation is shown in Fig. \ref{fig_el_struc} (top).
There are two parabolic bands  at special points ${\bf K}$ and ${\bf
  K^{\prime}}=2{\bf K}$ of the Brillouin zone (BZ). The
degree of trigonal warping is low and it increases by increasing doping,
as shown in the Fermi surface projection on the $k_x,k_y$ plane
in Fig. \ref{fig_FS}. 
The effective mass of the band, in the generalized gradient approximation (GGA)
within the PBE parametrization\cite{PBE} , 
is very close in the
two
compounds and weakly doping dependent. More details are given
in Refs.\cite{Pamuk2016,Pamuk2017}.

The static HF exchange interaction can be included in the electronic
structure calculation by using different
flavour of hybrid functionals, as shown in Fig. \ref{fig_el_struc}
(bottom). Beside increasing the gap between valence and conduction
band, the main effect of the exchange interaction is to slightly reduce the
effective
mass of the band ($m^*$).

\subsection{Vibrational properties and electron-phonon interaction}

\begin{figure*}[h]
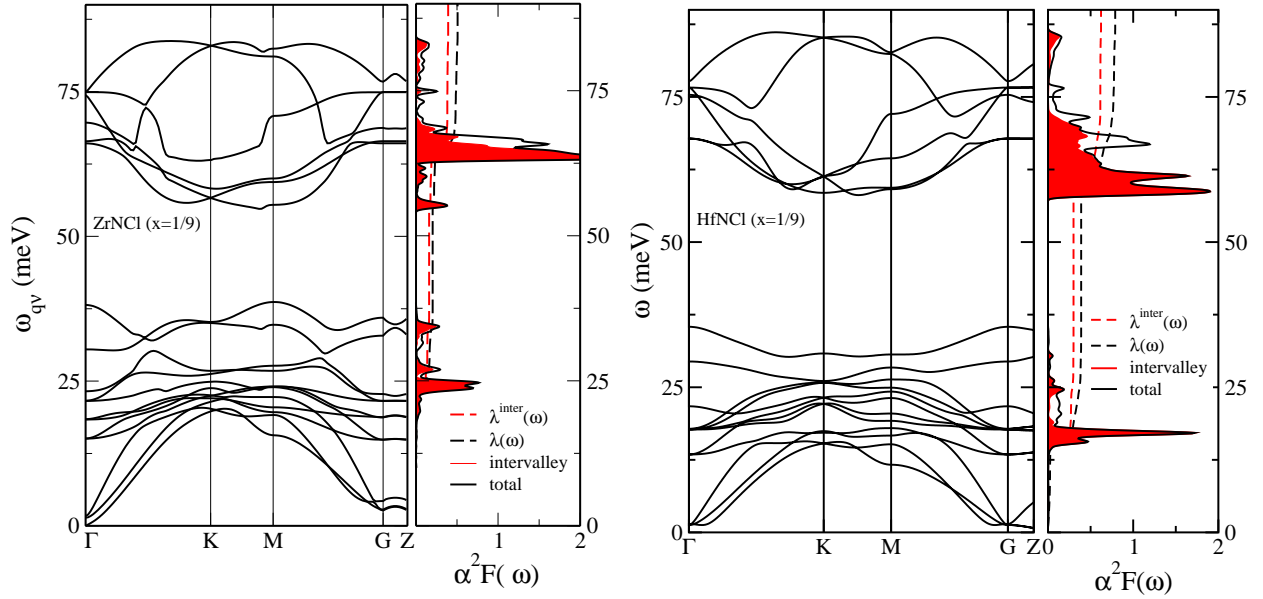

\vspace{1.0cm}
\centerline{\includegraphics[width=8cm]{ZrNCl_Fig_branchie_011.eps}\hspace{0.5cm}\includegraphics[width=8.25cm]{phonon_011_HfNCl_Betul.eps}}
\caption{(Color online) Phonon dispersion, Eliashberg function $\alpha^2F(\omega)$ 
and integrated Eliashberg function $\lambda(\omega)$ for Li$_x$(Zr,Hf)NCl at 
$x=1/9$. The intervalley contribution to $\alpha^2F(\omega)$ and $\lambda(\omega)$ are also shown.}
\label{fig_ph}
\end{figure*}

The phonon dispersions of Li$_x$(Zr,Hf)NCl at the doping fraction of $x=2/9$
calculated using Wannier interpolation as in Ref. \cite{WannierInterpol} are
shown in Fig. \ref{fig_ph}.
The high energy modes are very
similar in the two compounds, as shown in Fig. \ref{fig_ph}. These
modes are 
composed of Nitrogen vibrations and are separated from all the others.
Thus they are weakly affected by the replacement of Zr with Hf.
(Hf, Zr) and Cl modes are located in the low energy region below $40$
meV. Hf modes are softened due to the larger Hf mass.

The electron-phonon coupling of a given mode $\nu$
at a phonon-momentum \textbf{q} reads:
\begin{equation}
{\tilde{\lambda}}_{\mathbf{q}\nu}=\frac{2}{\omega_{\mathbf{q}\nu}^2 N(0) N_k} \sum_k |\tilde{d}^\nu_{\mathbf{k,k+q}}|^2\delta(\epsilon_\mathbf{k})\delta(\epsilon_{\mathbf{k+q}}),
\label{eq:lambda}
\end{equation}
where $\epsilon_{\mathbf{k}}$ is the quasiparticle energy of the
partially occupied band.
The electron-phonon matrix elements are defined such that
$\tilde{d}^\nu_{\mathbf{k,k+q}}=\langle\mathbf{k}|\delta\tilde{V}/\delta u_{\mathbf{q}\nu}|\mathbf{k+q}\rangle$,
$u_{\mathbf{q}\nu}$ is the phonon displacement of the mode $\omega_{\mathbf{q}\nu}$,
and $\tilde{V}$ is the single particle potential that is fully screened
by charge, spin, and valley exchange and correlation effects (see
Eq. 2 in Ref. \cite{Paolo2015} for more details). In the standard
implementation
of GGA/LDA functionals, there is no screening due to valley exchange
effects and $\tilde{V}$  is simply the Kohn-Sham potential
calculated with this approximation ($\tilde{V}=V_{KS}$).

The contribution of each mode to the electron-phonon interaction is
better
understood by looking at the Eliashberg function, namely:
\begin{equation}
\tilde{\alpha}^2F(\omega)=
\frac{1}{2 N_q}\sum_{{\bf q}\nu} {\tilde{\lambda}}_{{\bf q}\nu} \omega_{{\bf q}\nu} \delta(\omega-\omega_{{\bf q}\nu} )\label{eq:Elias}
\end{equation}
and the integral ${\tilde{\lambda}}(\omega)=2 \int_{0}^{\omega} d\omega^{\prime}
\tilde{\alpha}^2F(\omega^{\prime})/\omega^{\prime}$. 

We first calculate the electron-phonon interaction and the Eliashberg
function using the GGA approximation. Then we decompose the 
electron-phonon coupling into intravalley and intervalley
contributions. This can be
easily done as the two valleys are located in different regions
of the Brillouin zone and are disjoined. It is then sufficient to
select the proper electron and phonon momenta involving only intra or
intervalley scattering. The same decomposition is carried out on the
Eliashberg
function via Eq. \ref{eq:Elias}.

At the GGA level, the Eliashberg function is composed of two main
peaks, namely one at high energy  due to N vibrations and mostly
related to intervalley phonons (although a intravalley component is
present) and a second one at lower energy mostly of intervalley
character. The magnitude of the integrated total and intervalley 
electron-phonon couplings averaged
over the Brillouin zone are shown in Fig. \ref{fig_ph}. It is
clear that the electron-phonon interaction is dominated by intervalley
phonons. 
\begin{figure}[h]
\includegraphics[width=0.6\columnwidth]{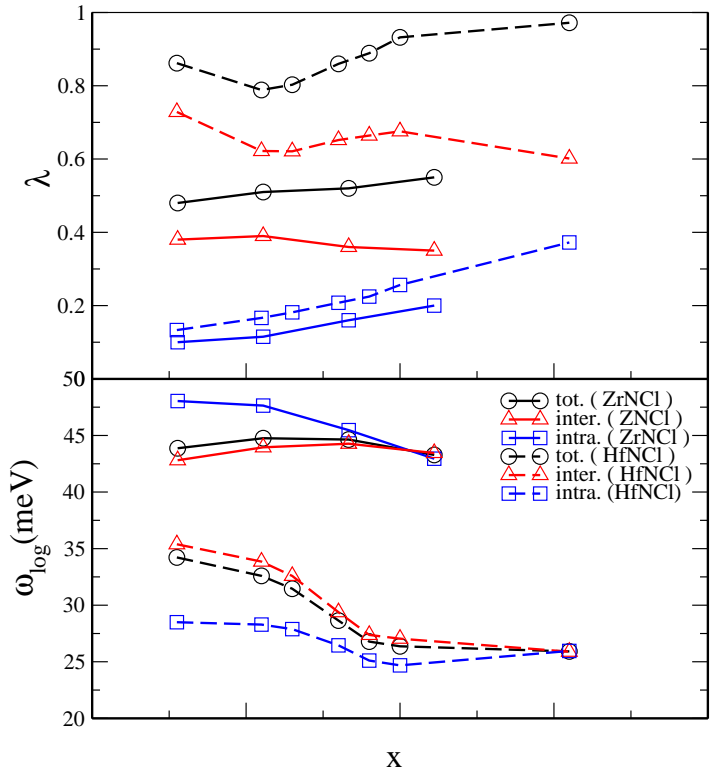}
\caption{(Color online) Electron-phonon coupling and logarithmic average of the
  phonon frequencies as a function of doping decomposed in the
  intervalley and intravalley contributions.}
\label{el_ph_pbe}
\end{figure}
The doping dependence of the electron phonon coupling  is represented
for the two systems in Fig. \ref{el_ph_pbe}. The total electron-phonon
coupling
is almost twice as large in Li$_x$HfNCl than in Li$_x$ZrNCl. This is
mostly due to the increased intervalley electron-phonon interaction
for the former compound. The intravalley coupling is roughly the same in both
compounds.

As the band structures for the two compounds are very similar,
the enhancement of the intervalley electron-phonon coupling is not due
to a band structure effect, but it is partly due the larger deformation potential for intervalley phonons
in  HfNCl and partly due to the softening of the main peaks of the
intervalley Eliashberg function in HfNCl and the consequent dependence
$\omega^{-2}$ in $\lambda^{\rm inter}$.

\begin{figure}
\centerline{\includegraphics[width=1.2\columnwidth]{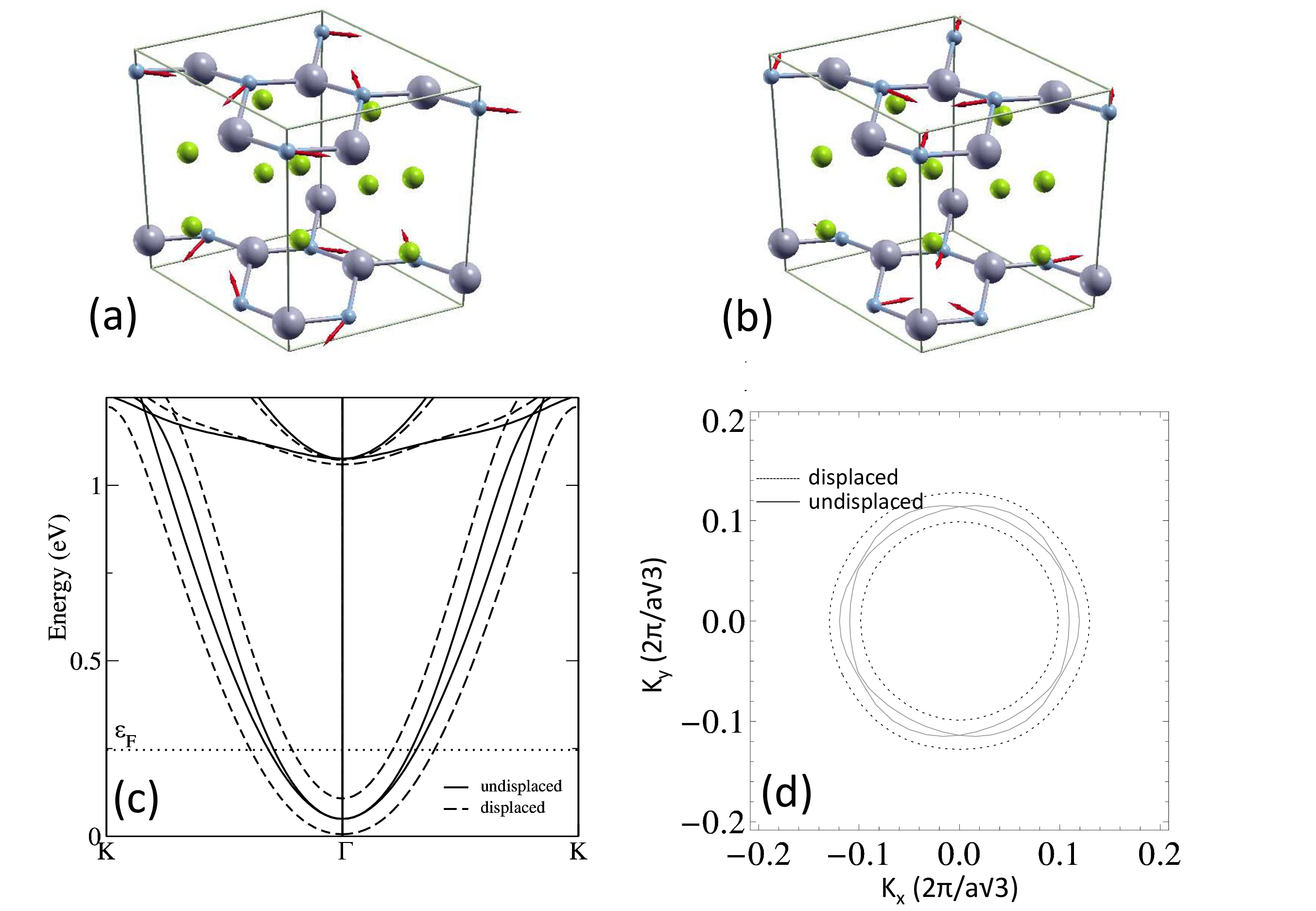}}
\caption{(Color online) $(a,b)$ Phonon pattern of the twofold degenerate intervalley phonon
  responsible for one of the prominent high-energy intervalley peaks in
  the Eliashberg function. The pattern is plotted in a 
$\sqrt{3}\times\sqrt{3}$ supercell. (Hf,Zr) big grey balls, Cl greenish balls and 
N cyan balls with arrows. The effect of the phonon patterns on the ZrNCl 
electronic structure $(c)$ and Fermi surface $(d)$ is shown in the bottom
panels $(c, d)$.  The two valleys at ${\bf K}$ and ${\bf K^{\prime}}$ in
the Brillouin zone of the unit cell  are folded at ${\bf \Gamma}$ in the
Brillouin
zone of the supercell. Intervalley phonons reduce the occupation of
one valley and increase the occupation of the other one, acting as a
pseudo-magnetic field on the valley degrees of freedom.
Phonon displacements associated to other intervalley vibrations
contributing to the electron-phonon interaction display an
analogous behaviour.}
\label{fig_patterns}
\end{figure}
Given the dominant role of the intervalley electron-phonon 
interaction, it is interesting to show the displacement pattern 
of these phonon modes 
and its effect on the 
electronic structure. For this reason, we consider a
$\sqrt{3}\times\sqrt{3}$ supercell and we displace the atoms according
to the patterns of the intervalley phonons generating the most
prominent peaks in the intervalley Eliashberg function. We then
calculate the electronic structure and Fermi surface for the displaced
and undisplaced atomic configurations. There are
several intervalley phonon modes contributing to $\alpha^2F(\omega)$, both at low energy
and high energy. As they have qualitatively
similar effects on the electronic structure, we just show in
Fig. \ref{fig_patterns} (a,b) the
action of the twofold degenerate high-energy mode mostly involving N
vibrations.

The effect of intervalley vibrations on the electronic structure is to
induce an unbalance in the occupation of the two valleys, as shown in
Fig.\ref{fig_patterns}. This means that each intervalley phonon acts
as  a pseudo magnetic field on the valley degrees of
freedom. In more details, if we assume 
 a constant intravalley electron-phonon matrix element
 ($|d_{{\bf k},{\bf k+K}}^{\nu}|\approx |d_{{\bf K},{\bf
     2K}}^{\nu}|$), the action of a small  phonon
displacement $u_{{\bf K}\nu}$ on the electronic structure can be
described by the following one body Hamiltonian 
in the basis formed by the  2-component spinors
 $|{\bf K}+\bm{\kappa}\rangle$ and $|2{\bf K}+\bm{\kappa}\rangle$,
with $\bm{\kappa}={\bf k}-{\bf K}$:
\begin{equation}
H^{\nu}_{\bm{\kappa}}=\frac{\hbar^2 \kappa^2}{2 m^*} \hat{I} + B_{\rm ext}^{\nu} \,\mu_S \hat{\sigma_x},
\end{equation}
where  $\hat{I}$ and $\hat{\sigma_x}$ are the $2\times2$ identity and the Pauli
 matrix along the x-direction, respectively, $B_{\rm ext}^{\nu} = |d_{{\bf K},{\bf 2K}}^{\nu}|u_{{\bf K}\nu}/\mu_S$

As it happens in the magnetic case for $\chi_{s}$, the
response to the pseudo magnetic field, $\chi_v$,  is enhanced by the
intervalley exchange-correlation, as predicted by Eq. \ref{eqsymchis}.
This effect is however absent in our DFT calculation, due to the lack
of dependence of the exchange correlation functional on intervalley
densities (see Supplemental of \cite{Paolo2015} for more details),
but it is normally present in hybrid-functional calculations via the
dependence of the exact exchange on electronic wavefunctions\cite{Pamuk2016,Pamuk2017}.

As the total magnetization due to the pseudo magnetic
field $B_{\rm ext}^{\nu}$ is written either as $M=\chi_s B_{\rm
  ext}^{\nu}$ or as $ M=\chi_{0s} \widetilde{B}^{\nu}$, where now $\widetilde{B}^{\nu}$
is the total magnetic field,
sum of the external plus the exchange-correlation field, we have,
\begin{eqnarray}
\frac{\widetilde{B}^{\nu}}{B_{\rm ext}^{\nu}}=
\frac{|\widetilde{d}_{{\bf K},{\bf 2K}}^{\nu}|}{|d_{{\bf K},{\bf
 2K}}^{\nu}|}=\frac{\chi_s}{\chi_{0s}}
\label{eq:drenK}
\end{eqnarray}
namely the electron-phonon coupling at ${\bf q}={\bf K}$ is
renormalized by the electron-electron interaction
exactly in the same way as the spin susceptibility with an enhancement that is independent from the phonon index $\nu$.
Assuming again a constant intervalley matrix element we have that:
\begin{eqnarray}
{\tilde \lambda}^{\rm inter}= \left(\frac{\chi_s}{\chi_{0s}}\right)^2
\lambda^{\rm inter}
\label{eq:lamtinter}
\end{eqnarray}
so that ${\tilde \lambda}=\lambda^{\rm intra}+{\tilde \lambda}^{\rm
  inter}$.
Thus the total electron-phonon interaction is enhanced in a way that
is proportional to the ratio between the fully interacting
valley susceptibility and the non interacting one.
The occurrence of this effect in ZrNCl and HfNCl can be explicitly
verified by using hybrid functional calculations with different
components of exact exchange \cite{Pamuk2016,Pamuk2017}.
In the framework of our RPA model we can just renormalize the
intervalley
electron-phonon interaction with the ration of the $\chi_s/\chi_{0s}$ 
in Fig. \ref{figs_chi_marchi}.

\section{Conclusions}
\begin{figure}[h]
\includegraphics[width=0.8\columnwidth]{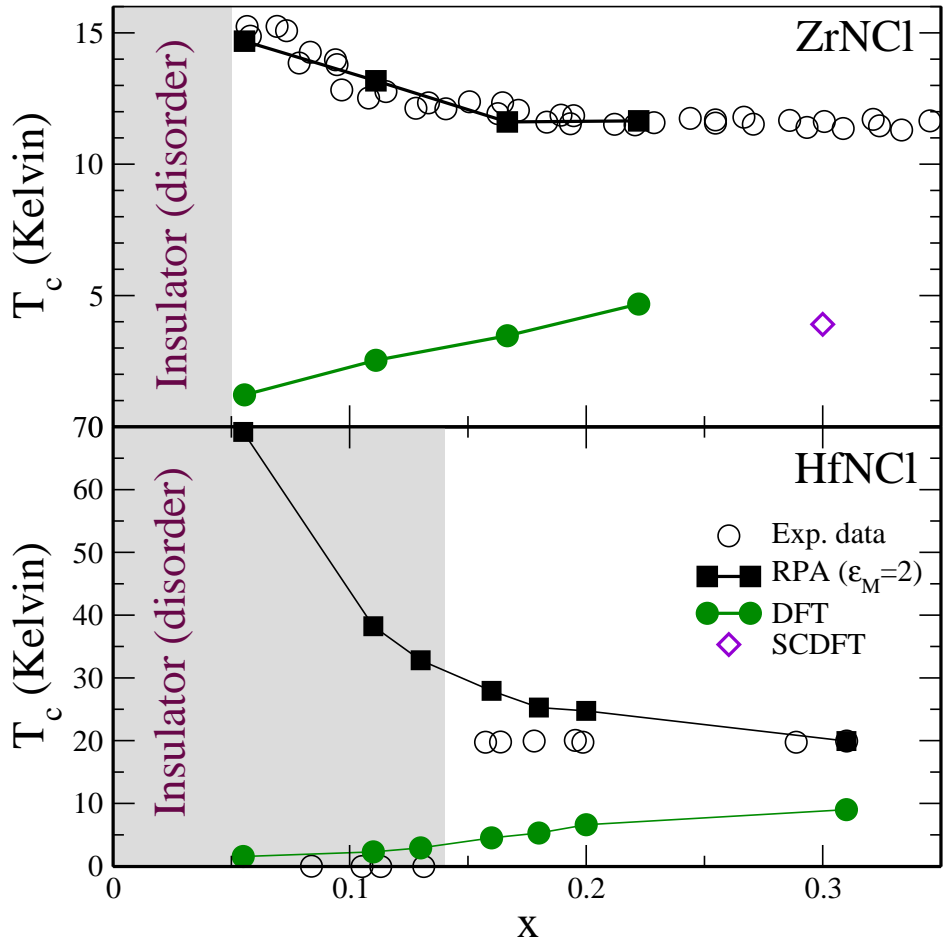}
\caption{(Color online) Superconducting critical temperatures calculated with
  different approximations and compared with experiments 
\cite{Kasahara2009,Takano2008}.
The grey region marks the occurrence of an insulating region detected
in experiments and due to disorder (Anderson transition).}
\label{fig_sc}
\end{figure}

The superconducting properties for both compounds as a function of
doping/intercalation are shown in Fig. \ref{fig_sc}. In experiments,
a metal insulator transition occurs at low doping due to the
occurrence of disorder. In intercalated HfNCl the Anderson insulating
state persists to doping as large as $x=0.14$, hindering the detection of
a T$_c$ enhancement in HfNCl. 

Density functional theory calculations based on gradient corrected
functionals lead to an estimate of T$_c$\cite{footnoteTc} that is too low compared to
experiments and it also has the incorrect behaviour of T$_c$ versus
doping. Indeed within the PBE parametrization \cite{PBE}, T$_c$ {\it increases} by {\it increasing }
doping.
Electron-electron interaction included within the RPA and via
renormalization of the intervalley electron-phonon interaction leads to
the correct T$_c$ versus doping behaviour and, for the case of
Li$_x$ZrNCl,
to an excellent agreement with experiments.
In HfNCl the comparison with experiments is complicated by the more persistent Anderson
transition and the difficulties of obtaining very good ordered samples
in the low doping limit.  
Our results predict that removal of the Anderson
transition or better control of doping in Li$_x$HfNCl
could lead to emergence of a high T$_c$ superconducting
state.


\section{Acknowledgments}

We acknowledge Stefania de Palo for providing us the data in
Fig. 1.
We acknowledge support from the acknowledge support from the European Union
Horizon 2020 research and innovation program under Grant
agreement No. 696656-GrapheneCore1 and from Agence
Nationale de la Recherche under the reference No. ANR-
13-IS10-0003-01. Computer
facilities provided by CINES, IDRIS, and CEA TGCC (Grant
EDARI No. 2017091202) and 
the institute for computing and data sciences (ISCD) at UPMC based in
France.
B. P. acknowledges National Science Foundation [Platform for the Accelerated
Realization, Analysis, and Discovery of Interface Materials (PARADIM)] under
Cooperative Agreement No. DMR-1539918 for her time at Cornell University.



\bibliographystyle{jpsj}
\bibliography{bibliography}

\end{document}